\documentclass[aps,prd,reprint,superscriptaddress,nofootinbib,floatfix,raggedbottom]{revtex4-2}
\usepackage{graphicx}
\usepackage{capt-of}
\usepackage{amsmath,amssymb}
\usepackage{booktabs,array}
\usepackage{microtype}
\usepackage{xcolor}
\usepackage[colorlinks=true,linkcolor=blue!40!black,citecolor=blue!40!black,urlcolor=blue!40!black]{hyperref}
\graphicspath{{./}}

\newcommand{\Mc}{M_\mathrm{c}}
\newcommand{\Mstar}{M_*}
\newcommand{\rc}{r_\mathrm{c}}
\newcommand{\Msun}{M_\odot}
\newcommand{\Gion}{\Gamma_{\mathrm{ion}}^{\mathrm{total}}}
\newcommand{\gion}{\Gamma_{\mathrm{ion}}}
\newcommand{\Pcgw}{P_{\mathrm{gw,c}}}
\newcommand{\dd}{\mathrm d}
\newcommand{\yr}{\,\mathrm{yr}}
\newcommand{\Myr}{\,\mathrm{Myr}}
\newcommand{\Gyr}{\,\mathrm{Gyr}}
\newcommand{\pc}{\,\mathrm{pc}}
\begin{document}
\title{Axion cloud survival in nuclear star clusters}
\author{Chen Feng}
\email{fengchen23@mails.ucas.ac.cn}
\affiliation{School of Astronomy and Space Science, University of Chinese Academy of Sciences (UCAS), Beijing 100049, China}

\begin{abstract}
Stellar perturbations can ionize a superradiant boson cloud while transferring energy and angular momentum from the perturbing orbits. We follow the cloud after superradiant saturation by coupling its occupation to a two-dimensional, orbit-averaged Fokker--Planck equation for the stars. The evolution includes stellar relaxation, capture, orbital backreaction, and gravitational-wave (GW) emission from both the cloud and the stellar orbits. We evolve 24 models with black hole masses $M=10^5$--$10^7\Msun$ and couplings $0.04\leq\alpha\leq0.40$. Stars with semimajor axes of a few Bohr radii, $\rc=M/\alpha^2$, dominate ionization. Stellar replenishment from the outer cluster sustains ionization over hundreds of Myr, particularly for extended clouds. Of the 24 models, 21 lose at least $99\%$ of their initial cloud mass. For compact clouds, intrinsic GW emission dominates the depletion.
\end{abstract}
\maketitle
\setlength{\jot}{2pt}
\setlength{\abovedisplayskip}{6pt plus 2pt minus 2pt}
\setlength{\belowdisplayskip}{6pt plus 2pt minus 2pt}
\setlength{\abovedisplayshortskip}{4pt plus 2pt}
\setlength{\belowdisplayshortskip}{4pt plus 2pt minus 2pt}
\setlength{\textfloatsep}{8pt plus 2pt minus 2pt}

\section{Introduction}\label{sec:introduction}
Ultralight bosons arise in extensions of the Standard Model and are candidates for dark matter \cite{Peccei:1977hh,Weinberg:1977ma,Wilczek:1977pj,Preskill:1982cy,Abbott:1982af,Dine:1982ah,Arvanitaki:2009fg,Marsh:2015xka,Hui:2016ltb,Hui:2021tkt}. Around a rotating black hole, superradiant instability can amplify gravitationally bound modes into a macroscopic cloud \cite{Zeldovich:1971ffh,Press:1972zz,Detweiler:1980uk,Dolan:2007mj,Pani:2012vp,Baryakhtar:2017ngi,East:2017ovw,East:2017mrj}. The resulting gravitational atom (GA) has a hydrogenic spectrum in the nonrelativistic limit. Cloud GW emission and the extraction of black hole spin provide observational probes of these fields \cite{Brito:2015oca,Baumann:2019eav,Arvanitaki:2010sy,Arvanitaki:2014wva,Arvanitaki:2016qwi,Brito:2017zvb,Brito:2017wnc,Stott:2018opm,Isi:2018pzk,Tsukada:2018mbp,Palomba:2019vxe}. Their interpretation depends on how long the cloud survives after superradiant growth saturates.

In a galactic nucleus, orbiting stars and compact remnants perturb the cloud. Their time-dependent gravitational potential can excite bound bosons into continuum states, producing ionization \cite{Baumann:2021fkf,Tomaselli:2023ysb,Cheng:2026dsq}. The associated transfer of energy and angular momentum changes the stellar orbits and hence their subsequent ionization rates \cite{Baumann:2022pkl,Cao:2023fyv,Cao:2024wby}. The cloud also depletes through intrinsic GW emission \cite{Brito:2014wla,Yoshino:2013ofa}. Bound-state transitions \cite{Baumann:2018vus,Baumann:2019ztm,Zhang:2019eid,Zhang:2018kib,Berti:2019wnn,Boskovic:2024fga,Tomaselli:2024dbw,Tomaselli:2024bdd,Tomaselli:2025jfo} and self-interactions \cite{Yoshino:2012kn,Baryakhtar:2020gao,Takahashi:2024fyq} can further modify the evolution; we restrict attention to ionization and intrinsic GW losses.

The supply of perturbing stars evolves with the surrounding cluster. Environmental effects on orbital motion and GW signals have been studied in several settings \cite{Sadeghian:2013laa,Eda:2013gg,Barausse:2014tra,Annulli:2020ilw,Annulli:2020lyc,Brito:2023pyl,Duque:2023seg}, while stellar and pulsar observations near the Galactic center probe the matter distribution \cite{GRAVITY:2018ofz,GRAVITY:2020gka,GRAVITY:2021xju,Yu:2025apk,Tomaselli:2025zdo}. Within a nuclear star cluster \cite{Neumayer:2020gno,Alexander:2017rvg}, gravitational encounters redistribute stellar energy and angular momentum, and orbital GW emission drives stars toward capture \cite{Bahcall:1976aa,Lightman:1977zz,Hopman:2005vr,Merritt:2015vxa}. These processes regulate the supply of extreme mass-ratio inspirals \cite{Gair:2004iv,Merritt:2011ve,Amaro-Seoane:2007osp,Amaro-Seoane:2012lgq,Babak:2017tow,LISA:2022yao} and also determine how rapidly depleted ionizing orbits are replenished.

Cloud depletion and stellar transport are therefore coupled. The evolving stellar distribution sets the total ionization rate, while the remaining cloud mass controls the strength of orbital backreaction. Relaxation can supply ionizing stars after the initially nearby population has been removed. Following this supply is essential for determining whether ionization shuts off or persists over the cloud's lifetime.

We follow this coupled evolution with a two-dimensional, orbit-averaged Fokker--Planck (FP) equation in stellar energy and angular momentum. The model includes two-body relaxation, orbital GW losses, capture, ionization backreaction, and intrinsic cloud GW emission. We consider a scalar $|211\rangle$ cloud in a single-mass Bahcall--Wolf cusp, allowing the black hole to grow through stellar capture. By resolving the ionizing orbits and tracking stars supplied by the outer cluster, we relate the cloud's mass history to the changing stellar population. We then compare the losses through ionization and intrinsic GWs with their effects on the depletion time.

Section~\ref{sec:model} presents our model for the coupled evolution of the cloud and stars. Section~\ref{sec:results} gives the numerical results for ionizing orbits, stellar replenishment, and cloud depletion, and Section~\ref{sec:conclusions} summarizes our conclusions. Appendices~\ref{app:ionization} and~\ref{app:stellar-dynamics} provide details of the single-orbit response and stellar diffusion and capture.

\section{Coupled cloud and stellar evolution}\label{sec:model}

\subsection{Cloud response}\label{subsec:axion-cloud}
We use $G=c=\hbar=1$ and denote the black hole mass and scalar particle mass by $M$ and $\mu$. The coupling $\alpha=M\mu$ controls the hydrogenic spectrum in the nonrelativistic limit \cite{Brito:2015oca}. We label the bound states $|nlm\rangle$, where $n,l,m$ are the principal, orbital, and azimuthal quantum numbers. The cloud occupies the co-rotating state $|1\rangle\equiv|211\rangle$, with occupation $N_{211}$ \cite{Baumann:2019eav}. Its Bohr radius, frequency, and mass are
\begin{equation}
 \rc=\frac{M}{\alpha^2},\quad
 \omega_1\approx\mu\left(1-\frac{\alpha^2}{8}\right),\quad \Mc=\mu N_{211}.
 \label{eq:cloud_scales}
\end{equation}
We write $\beta=\omega_1N_{211}/M$. Ionization requires an energy of at least $\mu-\omega_1$ per boson.

For a star of mass $\Mstar$ on an orbit with semimajor axis $a$ and eccentricity $e$, the mean frequency is $\Omega=\sqrt{M/a^3}$, the total orbital energy is $E_*=-M\Mstar/(2a)$, and the angular-momentum magnitude is $L=\Mstar\sqrt{Ma(1-e^2)}$. The ionization rate per boson is \cite{Baumann:2021fkf,Tomaselli:2023ysb}
\begin{equation}
 \gion(a,e)=\sum_{\substack{l,m,g\\k_g>0}} W_{glm},\qquad
 W_{glm}=\frac{\mu}{k_g}|\eta_{k_g lm}^{(g)}|^2.
 \label{eq:single_rate}
\end{equation}
Here $g$ is an integer orbital harmonic and $k_g=\sqrt{2\mu(\omega_1-\mu+g\Omega)}$ is the escaping boson's momentum. The continuum states $|k;lm\rangle$ have $k>0$, $l=0,1,\ldots$, and $-l\leq m\leq l$. The Fourier coefficient $\eta^{(g)}$ connects the bound state to these momentum-normalized states (Appendix~\ref{app:ionization}). Only harmonics with $g\Omega>\mu-\omega_1$ contribute. Each transition transfers energy $g\Omega$ and axial angular momentum $m-1$ from the star to the escaping boson, giving
\begin{align}
 P_{\mathrm{ion}}&=N_{211}\sum_{\substack{l,m,g\\k_g>0}}g\Omega W_{glm},\nonumber\\
 \tau_{{\mathrm{ion}},z}&=N_{211}\sum_{\substack{l,m,g\\k_g>0}}(m-1)W_{glm}.
 \label{eq:ion_exchange}
\end{align}
We evaluate the response for prograde equatorial orbits, with $\mathbf{L}=L\mathbf{e}_z$ and $\mathbf{e}_z$ along the cloud angular momentum. The orbital drifts are $\dot E_{*,\mathrm{ion}}=-P_{\mathrm{ion}}$ and $\dot L_{\mathrm{ion}}=-\tau_{{\mathrm{ion}},z}$; both scale with $N_{211}$. Averaging over independent, uniform initial mean anomalies removes interference between different stars (Appendix~\ref{app:ionization}). Let $n_*(E_*,L,t)\,\dd E_*\dd L$ be the number of stars in an orbital interval at time $t$, and let $\Pcgw$ be the intrinsic cloud GW power. Summing over the stellar population gives
\begin{align}
 \Gion(t)&=\int n_*(E_*,L,t)\gion(E_*,L)\,\dd E_*\dd L,\nonumber\\
 \dot N_{211}&=-N_{211}\Gion-\Pcgw/\omega_1.
 \label{eq:cloud}
\end{align}
The integral covers $a\leq r_{\mathrm{h}}$ and $L>L_{\mathrm{lc}}$, where $r_{\mathrm{h}}$ is the black hole influence radius and $L_{\mathrm{lc}}=4M\Mstar$ is the adopted capture threshold.

We use SuperRad, a Python package for modeling GW signals from superradiant boson clouds, to obtain the dimensionless intrinsic GW power $\widetilde P_{\mathrm{gw,c}}^{211}$ at saturation \cite{Siemonsen:2022yyf,Brito:2014wla,Yoshino:2013ofa,Brito:2017zvb},
\begin{equation}
 \Pcgw=\beta^2
 \widetilde P_{\mathrm{gw,c}}^{211}(\alpha).
 \label{eq:intrinsic_power}
\end{equation}

\subsection{Stellar distribution and initial conditions}\label{subsec:initial-conditions}
The stellar distribution evolves through energy and angular-momentum currents $F_E,F_L$ in an orbit-averaged FP description \cite{Merritt:2015vxa,Bar-Or:2015plb}. For coordinates $x_i\in\{E_*,L\}$, write $\partial_i=\partial/\partial x_i$ and sum repeated coordinate indices. The encounter drift and diffusion moments are $A_i=\langle\Delta x_i\rangle/\Delta t$ and $D_{ij}=\langle\Delta x_i\Delta x_j\rangle/\Delta t$ \cite{Chandrasekhar:1943ys,Rosenbluth:1957zz}, evaluated from the evolving scattering background (Appendix~\ref{app:stellar-dynamics}). With ionization drifts $\dot x_{i,\mathrm{ion}}$ and orbital GW drifts $\dot x_{i,\mathrm{GW}}$ from Peters \cite{Peters:1963ux,Peters:1964zz}, the equation reads
\begin{align}
 \partial_t n_*&=-\partial_{E_*}F_E-\partial_LF_L,\nonumber\\
 F_i&=(A_i+\dot x_{i,\mathrm{GW}}+\dot x_{i,\mathrm{ion}})n_*
 -\tfrac12\partial_j(D_{ij}n_*).
 \label{eq:fp}
\end{align}

After superradiant saturation, we set $\Mc(0)=0.01M(0)$ \cite{Yoshino:2012kn,Baryakhtar:2020gao}. The stars are identical compact objects of mass $\Mstar=10\Msun$, with mass ratio $q=\Mstar/M\ll1$. At distance $r$ from the black hole, their initial parent number density is the Bahcall--Wolf cusp \cite{Bahcall:1976aa}, $n_{\mathrm{iso}}(r)=n_{\mathrm{h}}(r/r_{\mathrm{h}})^{-\gamma_*}$, with slope $\gamma_*=7/4$ and normalization $n_{\mathrm{h}}=n_{\mathrm{iso}}(r_{\mathrm{h}})$. The one-dimensional stellar velocity dispersion $\sigma$ fixes the initial cusp scale through \cite{Ferrarese:2000se,Tremaine:2002js,Bar-Or:2015plb}
\begin{align}
 M&=5.4\times10^6\Msun\left(\frac{\sigma}{100\,\mathrm{km\,s^{-1}}}\right)^4,\nonumber\\
 r_{\mathrm{h}}&=\frac M{\sigma^2},\qquad n_{\mathrm{h}}=\frac{(3-\gamma_*)2M}{4\pi\Mstar r_{\mathrm{h}}^3}.
 \label{eq:cusp_normalization}
\end{align}
Here $M$ is evaluated at its initial value, so the parent stellar mass inside $r_{\mathrm{h}}$ is $2M(0)$. The outer radius and reservoir profile retain these initial values during the evolution. Let $T=2\pi/\Omega$ be the orbital period and $L_{\mathrm{c}}=\Mstar\sqrt{Ma}$ the circular-orbit angular momentum. The phase-space density $f$ is per $\dd^3r\dd^3p$, with momentum $\mathbf{p}=\Mstar\mathbf{v}$ and stellar velocity $\mathbf{v}$. Its initial value is the isotropic parent distribution $f_{\mathrm{iso}}$ multiplied by a logarithmic loss-cone factor \cite{Merritt:2015vxa}:
\begin{equation}
 f(E_*,L,0)=f_{\mathrm{iso}}(E_*)
 \frac{\ln(L^2/L_{\mathrm{lc}}^2)}{\ln(L_{\mathrm{c}}^2/L_{\mathrm{lc}}^2)},\quad
 n_*=8\pi^2TLf.
 \label{eq:initial_shape}
\end{equation}
We set $f=0$ below $L_{\mathrm{lc}}$. The factor $8\pi^2TL$ converts $f$ to the orbital number density $n_*$. Appendix~\ref{app:stellar-dynamics} gives the normalization of $f_{\mathrm{iso}}\propto(-E_*)^{1/4}$.

At the fixed outer boundary $a=r_{\mathrm{h}}$, we hold $n_*$ at its initial value. At $L=L_{\mathrm{c}}(E_*)$, the condition $F_L-(\dd L_{\mathrm{c}}/\dd E_*)F_E=0$ forbids flux across the circular-orbit boundary. We use the Cohn--Kulsrud encounter fluxes at $L=L_{\mathrm{lc}}$ \cite{Merritt:2015vxa}; stars crossing this boundary under orbital GW or ionization drifts are also removed.

We assume that stellar orbital orientations remain isotropically distributed throughout the evolution, neglecting orientation-dependent depletion by ionization. The stellar FP equation uses a spherical scattering background, and the present ionization response is evaluated for prograde equatorial Keplerian orbits. We consider initial black hole masses $M(0)=10^5,10^6,10^7\Msun$ and the eight initial couplings in Table~\ref{tab:scan}. We include black hole growth from stellar capture at fixed $\mu$ and stop each run after at least $99\%$ cloud depletion, at $10\Gyr$, or at $\alpha=0.45$, which we adopt as the limit of validity of our approximations. As the black hole slowly gains mass, stellar orbits shrink, and we include this effect when evolving the stellar distribution.

\begin{figure*}[t]
 \centering
 \includegraphics[width=\textwidth]{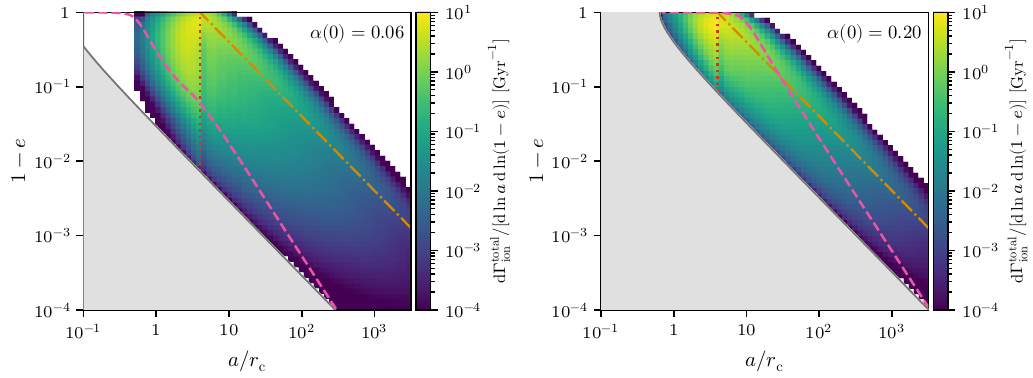}
 \caption{Orbital contributions at peak stellar ionization for initial black hole mass $M(0)=10^6\Msun$: $\alpha(0)=0.06$ (left) and $0.20$ (right). Both models include black hole growth and intrinsic cloud GWs. Colors give the rate per logarithmic orbital interval, with a common scale. Magenta dashed curves show Eq.~\eqref{eq:times}; grey marks capture. The red dotted and gold dash-dotted guides mark $a=4\rc$ and $a(1-e)=4\rc$, respectively. The cloud scale and all curves use the parameters at the corresponding peak.}
 \label{fig:phase}
\end{figure*}

The cloud is assumed to remain close to saturation as the black hole grows through stellar capture. For our representative models with $M(0)=10^6\Msun$ and $\alpha(0)=0.06$ and $0.20$, the shortest capture-driven mass-growth timescales $M/\dot M$ are approximately $1.8$ and $1.7\Gyr$, respectively. For comparison, SuperRad gives cloud-mass e-folding times of approximately $0.8\Myr$ and $50\yr$ at these initial masses and couplings for a dimensionless black hole spin of $0.9$ \cite{Siemonsen:2022yyf}. These estimates provide qualitative support for the adopted approximation in these representative cases.

Moreover, stellar capture is isotropic in our model and adds no net angular momentum on average. As the black hole grows, it can absorb part of the cloud to gain the angular momentum needed to restore saturation \cite{Hui:2022sri}. Cloud absorption would therefore provide an additional loss channel rather than lead to cloud growth. Under these assumptions, we neglect the additional cloud depletion associated with re-saturation, as this process is outside the scope of the present work.

We track the remaining fraction $S=\Mc(t)/\Mc(0)=N_{211}(t)/N_{211}(0)$ and define $t_{99}$ by $S(t_{99})=0.01$. Ionization and intrinsic GW losses are integrated separately along the coupled mass history.

We use 256 cells along each logarithmic coordinate, $\ln[a/(16M)]$ and $\ln(L^2/L_{\mathrm{lc}}^2)$, with a positive, conservative implicit discretization. Encounter coefficients are updated each step. Cell-averaged ionization responses resolve harmonic thresholds; the cloud equation is integrated analytically at fixed $\Gion$ within each split step. With black hole growth included, reference depletion times change by $0.43\%$ and $0.006\%$ between 192 and 256 cells for $\alpha(0)=0.06$ and $0.20$, respectively. Joint time-step and encounter-quadrature refinement changes the $\alpha(0)=0.06$ result by $0.14\%$ at fixed 128 cells per axis.

\section{Results}\label{sec:results}

\begin{figure}[t]
 \centering
 \includegraphics[width=\linewidth]{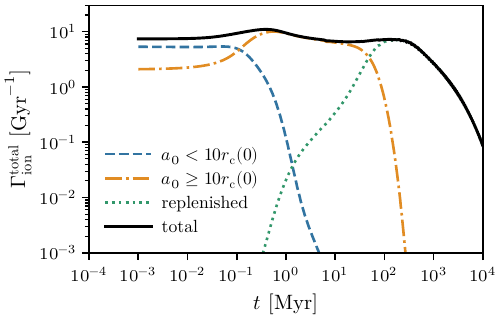}
 \caption{Ionization rate for $M(0)=10^6\Msun$ and $\alpha(0)=0.06$, including black hole growth and intrinsic cloud GWs. Dashed and dash-dotted curves separate the initial population at $a_0=10\rc(0)$, where $a_0=a(0)$. The dotted curve tracks stars supplied through the fixed outer boundary $a=r_{\mathrm{h}}$, including re-entry; the solid curve is their sum.}
 \label{fig:rate}
\end{figure}

\begin{figure*}[t]
 \centering
 \includegraphics[width=\textwidth]{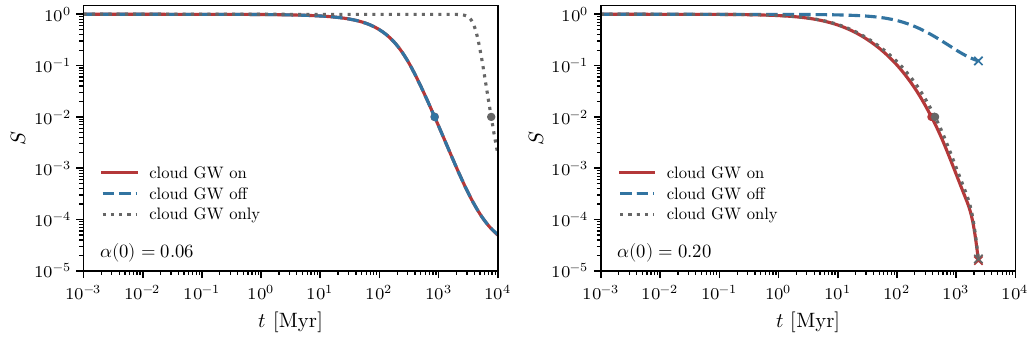}
 \caption{Remaining cloud fraction $S=N_{211}(t)/N_{211}(0)$ for $M(0)=10^6\Msun$, $\alpha(0)=0.06$ (left) and $0.20$ (right). ``cloud GW on'' is the coupled evolution with black hole growth. ``cloud GW off'' and ``cloud GW only'' remove the corresponding cloud-loss term while retaining the saved black hole mass and stellar histories; they do not recompute stellar feedback. Dots mark $S=0.01$. Curves end at the last available state; crosses in the right panel mark the coupling limit.}
 \label{fig:survival}
\end{figure*}

\subsection{Ionizing orbits and replenishment}

Figure~\ref{fig:phase} resolves the peak ionization rate in the $(a,1-e)$ plane for $M(0)=10^6\Msun$ and $\alpha(0)=0.06,0.20$. In both cases, stars with semimajor axes of a few $\rc$ dominate. We compare relaxation with orbital GW circularization using the diagnostic \cite{Hopman:2005vr}
\begin{equation}
 t_{\mathrm{rel}}|\dot e_{\mathrm{GW}}|=1,\qquad t_{\mathrm{rel}}=\frac{L_{\mathrm{c}}^2}{D_{LL}}.
 \label{eq:times}
\end{equation}
Here $t_{\mathrm{rel}}$ is the angular-momentum relaxation scale and $\dot e_{\mathrm{GW}}$ is the orbital GW eccentricity drift. Most of the peak rate lies at $t_{\mathrm{rel}}|\dot e_{\mathrm{GW}}|<1$ for $\alpha=0.06$ and at $t_{\mathrm{rel}}|\dot e_{\mathrm{GW}}|>1$ for $\alpha=0.20$. The extended cloud's ionizing orbits lie farther out, where encounters are more effective relative to orbital GW losses. This diagnostic concerns eccentricity drift: near circularity, $\dot e_{\mathrm{GW}}\to0$ while GW-driven orbital shrinkage remains finite.

Figure~\ref{fig:rate} shows how replenishment sustains ionization for $\alpha(0)=0.06$. We track the initial population in two groups, with initial semimajor axis $a_0<10\rc(0)$ and $a_0\geq10\rc(0)$, and a third group supplied through the boundary $a=r_{\mathrm{h}}$. The first peak, near $0.394\Myr$, is dominated by stars present initially. As their contribution declines, replenished stars produce a broad second maximum near $130\Myr$ and dominate the later rate. Thus the sustained ionization draws on stars supplied by the outer reservoir. The replenished component includes any stars that leave and subsequently re-enter the domain. At later times, the increase in $\alpha$ makes the cloud more compact and weakens stellar ionization.

\subsection{Cloud depletion}

Figure~\ref{fig:survival} shows the remaining cloud fraction with black hole growth. For $\alpha(0)=0.06$, the coupled evolution reaches $S=0.01$ at about $856\Myr$. The curve obtained by omitting intrinsic cloud GWs follows it closely during the main depletion stage. Stellar ionization accounts for almost all of the depletion in this case.

For $\alpha(0)=0.20$, intrinsic GWs dominate the mass loss, and the coupled evolution reaches $S=0.01$ at about $382\Myr$. With ionization omitted from the cloud equation but the same black hole mass history retained, this threshold is reached at about $441\Myr$. These channel comparisons use the saved evolution rather than separate stellar FP calculations. The right-panel curves stop at $\alpha\simeq0.45$.

\begin{table}[!t]
 \centering
 \caption{Models with initial parameters $M(0)$ and $\alpha(0)$, $\Mstar=10\Msun$, and $\Mc(0)=0.01M(0)$. The ionization and cloud GW columns give losses as percentages of the initial cloud mass. Times have two significant figures; a dash denotes that $S=0.01$ was not reached in the saved evolution.}
 \label{tab:scan}
 \small
 \setlength{\tabcolsep}{3pt}
 \renewcommand{\arraystretch}{0.90}
 \begin{tabular}{@{}ccccc@{}}
 \toprule
 $M(0)/\Msun$ & $\alpha(0)$ & \shortstack{$t_{99}$\\{[Myr]}}
 & \shortstack{Ionization\\{[\%]}}
 & \shortstack{Cloud GW\\{[\%]}}\\
 \midrule
$10^{5}$ & 0.04 & 28 & 100.00 & $6.29\times10^{-8}$ \\
$10^{5}$ & 0.06 & 32 & 100.00 & $2.27\times10^{-5}$ \\
$10^{5}$ & 0.10 & 47 & 99.95 & 0.0369 \\
$10^{5}$ & 0.15 & 69 & 88.61 & 11.20 \\
$10^{5}$ & 0.20 & 39 & 19.73 & 80.20 \\
$10^{5}$ & 0.25 & 9.2 & 2.25 & 97.74 \\
$10^{5}$ & 0.30 & 1.8 & 0.363 & 99.63 \\
$10^{5}$ & 0.40 & 0.097 & 0.0151 & 99.98 \\
\midrule
$10^{6}$ & 0.04 & 860 & 99.22 & $1.71\times10^{-7}$ \\
$10^{6}$ & 0.06 & 860 & 99.22 & $5.61\times10^{-5}$ \\
$10^{6}$ & 0.10 & 1,500 & 99.55 & 0.45 \\
$10^{6}$ & 0.15 & 1,100 & 65.56 & 34.44 \\
$10^{6}$ & 0.20 & 380 & 10.72 & 88.69 \\
$10^{6}$ & 0.25 & 89 & 1.27 & 98.67 \\
$10^{6}$ & 0.30 & 18 & 0.22 & 99.76 \\
$10^{6}$ & 0.40 & 0.96 & 0.0107 & 99.99 \\
\midrule
$10^{7}$ & 0.04 & \textemdash & 83.29 & $6.35\times10^{-5}$ \\
$10^{7}$ & 0.06 & \textemdash & 85.74 & 0.0114 \\
$10^{7}$ & 0.10 & \textemdash & 73.56 & 14.91 \\
$10^{7}$ & 0.15 & 6,700 & 30.64 & 69.17 \\
$10^{7}$ & 0.20 & 2,600 & 5.31 & 94.25 \\
$10^{7}$ & 0.25 & 750 & 0.728 & 99.10 \\
$10^{7}$ & 0.30 & 170 & 0.143 & 99.84 \\
$10^{7}$ & 0.40 & 9.6 & 0.00721 & 99.93 \\
 \bottomrule
 \end{tabular}
\end{table}

\begin{samepage}
Table~\ref{tab:scan} compares the depletion times and the two mass-loss channels across the initial parameters. Ionization dominates the mass loss at $\alpha(0)<0.15$, and intrinsic GWs at $\alpha(0)\geq0.20$.
\par
\end{samepage}

If the surrounding dark matter halo contains the same axions as the cloud, capture of dark matter axions could replenish the cloud and reduce its net depletion rate \cite{Budker:2023sex,Branco:2026mpt}. These cloud depletion histories have direct implications for the cloud's intrinsic GW signal \cite{Brito:2017zvb,Isi:2018pzk}. At fixed black hole parameters, the GW strain amplitude and luminosity scale as $\Mc$ and $\Mc^2$, respectively \cite{Arvanitaki:2014wva,Siemonsen:2022yyf}. Stellar ionization can therefore reduce the GW strain amplitude at a given cloud age and shorten the duration of the detectable GW signal, even when intrinsic GWs account for most of the cloud's mass loss.

\section{Conclusions}\label{sec:conclusions}
The survival of a scalar $|211\rangle$ cloud in a nuclear star cluster depends on the evolving supply of ionizing stars. We have followed this supply with a two-dimensional, orbit-averaged FP equation coupled to cloud depletion through ionization and intrinsic GW emission. Stellar relaxation, capture, orbital GW losses, and ionization backreaction jointly determine the population that remains available to perturb the cloud.

Stars with semimajor axes of a few Bohr radii dominate ionization. For the extended reference cloud, $M(0)=10^6\Msun$ and $\alpha(0)=0.06$, the rate first peaks as the initial stellar population evolves, then has a broad second maximum near $130\Myr$ as stars arrive from the outer reservoir. This replenishment sustains ionization over hundreds of Myr, allowing the cloud to reach $99\%$ depletion in about $856\Myr$ even after the initial contribution has declined.

Of the 24 models, 21 reach $99\%$ cloud depletion. Ionization dominates the integrated loss for $\alpha(0)<0.15$; intrinsic GWs dominate for $\alpha(0)\geq0.20$. The latter regime still depends on the stellar environment. For $M(0)=10^6\Msun$ and $\alpha(0)=0.20$, ionization removes only $10.7\%$ of the initial mass but shortens $t_{99}$ from $441\Myr$ to about $382\Myr$ along the same black hole mass history. The cumulative loss fractions alone therefore do not quantify the effect of either channel on the cloud's lifetime.

These depletion times apply to the adopted $10\Msun$ single-mass cusp, fixed outer reservoir, and prograde equatorial ionization response. The cloud response and stellar orbits are treated in the hydrogenic and Keplerian approximations, with the largest couplings representing nonrelativistic extrapolations. We do not include transitions
between bound cloud states. Within this model, stellar replenishment sustains the loss of extended clouds, while ionization accelerates the depletion of compact clouds even when intrinsic GWs remove most of their mass.

\begin{acknowledgments}
I thank Yan Cao for helpful discussions and valuable input, and Yong Tang for insightful guidance.
\end{acknowledgments}

\appendix
\section{Single-orbit response and drifts}\label{app:ionization}
The hydrogenic bound states describe the leading nonrelativistic scalar dynamics \cite{Zouros:1979iw,Dolan:2012yt,Witek:2012tr}. In the accelerating black-hole frame, write $\mathbf{r}$ for the boson position and $\mathbf{r}_*(t)$ for the stellar position, with $r=|\mathbf{r}|$, $r_*=|\mathbf{r}_*|$, and $\hat{\mathbf{r}}_*=\mathbf{r}_*/r_*$. The stellar potential energy is \cite{Baumann:2021fkf,Tomaselli:2023ysb}
\begin{equation}
 V_*=-\Mstar\mu\left(\frac1{|\mathbf{r}-\mathbf{r}_*|}
 -\frac{\mathbf{r}\cdot\mathbf{r}_*}{r_*^3}\right).
 \label{eq:perturbing_potential}
\end{equation}
The second term accounts for the acceleration of the black hole. Let $\theta,\phi$ be polar and azimuthal angles, with solid-angle element $\dd\Omega_2=\sin\theta\,\dd\theta\dd\phi$. Using unit-normalized spherical harmonics $Y_{lm}(\theta,\phi)$ and regular Coulomb functions $F_l$, the wavefunctions are $\psi_{211}=R_{21}Y_{11}$ and $\psi_{klm}=R_{kl}Y_{lm}$, with radial factors
\begin{align}
 R_{21}(r)&=\frac{r/\rc}{2\sqrt6\,\rc^{3/2}}e^{-r/(2\rc)},\nonumber\\
 R_{kl}(r)&=\frac2r F_l[-1/(k\rc),kr].
 \label{eq:wavefunctions}
\end{align}
The continuum frequency is $\omega(k)=\mu+k^2/(2\mu)$ for $k>0$. Momentum states have Dirac-delta normalization,
\begin{equation}
 \langle k;lm|k';l'm'\rangle=2\pi\delta(k-k')\delta_{ll'}\delta_{mm'}.
 \label{eq:continuum_normalization}
\end{equation}
The Kronecker deltas fix the angular quantum numbers. At the continuum threshold, the Bessel function of the first kind $J_{2l+1}$ gives
\begin{equation}
 \lim_{k\to0}\frac{R_{kl}(r)}{\sqrt{k}}
 =2\sqrt{\frac\pi r}\,J_{2l+1}(\sqrt{8r/\rc}).
 \label{eq:continuum_threshold}
\end{equation}
For integer harmonic $g$, the Fourier coefficient is
\begin{equation}
 \eta_{klm}^{(g)}=\frac1T\int_0^T\dd t\,e^{ig\Omega t}
 \int\dd^3r\,\psi_{klm}^*V_*\psi_{211}.
 \label{eq:fourier}
\end{equation}
For potential multipoles $l_*\geq0$ and $|m_*|\leq l_*$, expansion of the direct term at fixed $k,l$ gives the radial integral
\begin{align}
 \mathcal{I}_{l_*}(r_*)={}&\frac1{r_*^{l_*+1}}\int_0^{r_*}\!r^{l_*+2}R_{kl}^*R_{21}\,\dd r\nonumber\\
 &+r_*^{l_*}\int_{r_*}^{\infty}\!r^{1-l_*}R_{kl}^*R_{21}\,\dd r
 \label{eq:radial_integral}
\end{align}
and the instantaneous matrix element, with Gaunt coefficient $\mathcal G_{lm,l_*m_*}$,
\begin{align}
 \eta^{\mathrm{dir}}_{klm}(t)&=-4\pi\Mstar\mu\sum_{l_*,m_*}
 \frac{Y_{l_*m_*}^*(\hat{\mathbf{r}}_*)}{2l_*+1}\mathcal G_{lm,l_*m_*}\mathcal{I}_{l_*}(r_*),\nonumber\\
 \mathcal G_{lm,l_*m_*}&=\int Y_{lm}^*Y_{l_*m_*}Y_{11}\,\dd\Omega_2.
 \label{eq:angular_integral}
\end{align}
The angular integral requires $m_*=m-1$, $|l-1|\leq l_*\leq l+1$, and even $l+l_*+1$. The indirect term is included before taking the Fourier transform. First-order perturbation theory gives the rate \cite{Baumann:2021fkf,Tomaselli:2023ysb}
\begin{equation}
 \gion=\sum_{l,m,g}\int_0^\infty\!\dd k\,|\eta_{klm}^{(g)}|^2\delta[\omega(k)-\omega_1-g\Omega].
 \label{eq:golden_rule}
\end{equation}
Integration over $k$ produces the density-of-states factor $\mu/k_g$ in Eq.~\eqref{eq:single_rate}.

We retain $l_*\leq12$; doubling this cutoff changes the rate and exchange fluxes by less than $0.0006\%$ for the representative orbits tested. For $a>64\rc$, we replace the dense harmonic sum by an interpolation between the response at $a=64\rc$ and its parabolic limit. Discrete-sum checks near this boundary differ by up to $2.2\%$ for individual orbits; stars with $a>64\rc$ supply only $1.46$--$2.19\%$ of the reference peak rates.

For star label $s$, the initial mean anomaly $\mathcal M_{0,s}$ contributes a phase $e^{-ig\mathcal M_{0,s}}$. Averaging over independent uniform anomalies, with $s'$ another star label and $g'$ another harmonic, gives
\begin{equation}
 \left\langle e^{-ig\mathcal M_{0,s}}e^{ig'\mathcal M_{0,s'}}\right\rangle=
 \begin{cases}
 \delta_{g0}\delta_{g'0},&s\ne s',\\
 \delta_{gg'},&s=s'.
 \end{cases}
 \label{eq:phase_average}
\end{equation}
Since ionization requires $g>0$, different-star interference vanishes even for equal orbital frequencies. The escaping boson carries total energy $\omega_1+g\Omega$ and axial angular momentum $m$. Subtracting the bound boson's energy and angular momentum yields the orbital transfers used in Eq.~\eqref{eq:ion_exchange}.

For either dissipative mechanism, the orbital elements obey
\begin{equation}
 \dot a=\frac{2a^2\dot E_*}{M\Mstar},\qquad
 \frac{\dd e^2}{\dd t}=\frac{2L^2\dot E_*+4E_* L\dot L}{M^2\Mstar^3}.
 \label{eq:orbital_drifts}
\end{equation}
The $e^2$ form is regular at $e=0$. The leading orbital GW drifts for $q\ll1$ are \cite{Peters:1963ux,Peters:1964zz}
\begin{align}
 \dot a_{\mathrm{GW}}&=-\frac{64\Mstar M^2(1+73e^2/24+37e^4/96)}{5a^3(1-e^2)^{7/2}},\nonumber\\
 \dot e_{\mathrm{GW}}&=-\frac{304\Mstar M^2e(1+121e^2/304)}{15a^4(1-e^2)^{5/2}}.
 \label{eq:peters}
\end{align}
They enter the FP currents through
\begin{align}
 \dot E_{*,\mathrm{GW}}&=\frac{M\Mstar}{2a^2}\dot a_{\mathrm{GW}},\nonumber\\
 \dot L_{\mathrm{GW}}&=\frac L2\left(\frac{\dot a_{\mathrm{GW}}}a-\frac{2e\dot e_{\mathrm{GW}}}{1-e^2}\right).
 \label{eq:stellar_gw_conversion}
\end{align}
For $\alpha\geq0.30$, the hydrogenic response and Keplerian dynamics are nonrelativistic extrapolations.

\section{Stellar diffusion and capture}\label{app:stellar-dynamics}
The drift and diffusion moments describe encounter-driven changes in $x_i\in\{E_*,L\}$ \cite{Chandrasekhar:1943ys,Rosenbluth:1957zz}. Loss-cone transport determines how encounters supply stars to capture \cite{Frank:1976uy,Lightman:1977zz,Magorrian:1999vm,Wang:2003ny,Vasiliev:2017sbo}. We retain uncorrelated two-body relaxation, without mass segregation \cite{Alexander:2008tq,Preto:2009kd} or coherent torques \cite{Rauch:1996fb,Hopman:2006qr}.

The stellar energy and speed are related by $E_*(r,v)=\Mstar(v^2/2-M/r)$, with $v=|\mathbf{v}|$. The isotropic parent cusp satisfies \cite{Merritt:2015vxa}
\begin{equation}
 n_{\mathrm{iso}}(r)=4\pi\Mstar^3\int_0^{\sqrt{2M/r}}
 f_{\mathrm{iso}}[E_*(r,v)]v^2\,\dd v
 \label{eq:density_normalization}
\end{equation}
fixes the distribution
\begin{equation}
 f_{\mathrm{iso}}(E_*)=\frac{n_{\mathrm{h}}[r_{\mathrm{h}}/(M\Mstar)]^{\gamma_*}(-E_*)^{\gamma_*-3/2}}
 {4\pi\sqrt2\,\Mstar^{3/2}\mathrm{B}(3/2,\gamma_*-1/2)}.
 \label{eq:initial_distribution}
\end{equation}
Here $\mathrm{B}$ is the beta function; the $\Mstar^3$ factor converts velocity to canonical-momentum volume \cite{Merritt:2015vxa}. The reference model has $r_{\mathrm{h}}\approx1\pc$ and approximately $1.67\times10^5$ stars after the orbital cuts. For scattering-star speed $u$, the angular-momentum average $\bar f$ and local isotropized velocity distribution $f_{\mathrm{b}}$ are \cite{Merritt:2015vxa}
\begin{align}
 \bar f(E_*,t)&=\frac2{L_{\mathrm{c}}^2}\int_{L_{\mathrm{lc}}}^{L_{\mathrm{c}}}Lf\,\dd L,\nonumber\\
 f_{\mathrm{b}}(r,u,t)&=\Mstar^3\bar f[E_*(r,u),t].
 \label{eq:field_distribution}
\end{align}
For $a>r_{\mathrm{h}}$ the initial parent power law is fixed; no scattering stars are retained for $a<16M$. With $C=2(4\pi\Mstar)^2\ln(M/\Mstar)/3$ and $v^2=2(E_*/\Mstar+M/r)$, define the velocity moments $I_j$ ($j=2,4$) and $J$ at fixed $r,t$, suppressing these arguments:
\begin{equation}
 I_j=\int_0^v u^j f_{\mathrm{b}}(u)\,\dd u,\qquad
 J=\int_v^{\sqrt{2M/r}}u f_{\mathrm{b}}(u)\,\dd u.
 \label{eq:velocity_moments}
\end{equation}
The local velocity drift $a_\parallel$ and variances $b_\parallel,b_\perp$ are resolved along and perpendicular to $\mathbf{v}$ \cite{Chandrasekhar:1943ys,Rosenbluth:1957zz,Bar-Or:2015plb}:
\begin{align}
 a_\parallel&=-\frac{3CI_2}{v^2},\qquad
 b_\parallel=C\left(\frac{I_4}{v^3}+J\right),\nonumber\\
 b_\perp&=C\left(\frac{3I_2}v-\frac{I_4}{v^3}+2J\right),
 \label{eq:velocity_integrals}
\end{align}
where $b_\perp$ sums both perpendicular variances. The energy and angular-momentum moments are
\begin{align}
 A_E(r)&=\Mstar[va_\parallel+(b_\parallel+b_\perp)/2],\nonumber\\
 A_L(r)&=La_\parallel/v+\Mstar^2r^2b_\perp/(4L),\nonumber\\
 D_{EE}(r)&=\Mstar^2v^2b_\parallel,\quad D_{EL}(r)=D_{LE}(r)=\Mstar Lb_\parallel,\nonumber\\
 D_{LL}(r)&=\frac{L^2}{v^2}b_\parallel
 +\frac12\left(\Mstar^2r^2-\frac{L^2}{v^2}\right)b_\perp.
 \label{eq:encounter_moments}
\end{align}
Their time averages along each Kepler orbit enter Eq.~\eqref{eq:fp}.

Capture uses the near-parabolic threshold $L_{\mathrm{lc}}=4M\Mstar$, or $a(1-e^2)=16M$, intersecting circular orbits at $a=16M$. The Cohn--Kulsrud parameter in our variables is \cite{Merritt:2015vxa}
\begin{equation}
 q_{\mathrm{lc}}=\frac T2\left(\frac{D_{EE}}{E_*^2}+\frac{4D_{LL}}{L^2}
 +\frac{4D_{EL}}{E_* L}\right)_{L=L_{\mathrm{lc}}}.
 \label{eq:ck_parameter}
\end{equation}
Let $j_{0n}$ be the $n$th positive zero of $J_0$ and $f_{\mathrm{lc}}=f(E_*,L_{\mathrm{lc}},t)$. The boundary-layer function and capture gradient are \cite{Merritt:2015vxa}
\begin{align}
 \xi(q_{\mathrm{lc}})&=1-4\sum_{n=1}^{\infty}\frac{e^{-j_{0n}^2q_{\mathrm{lc}}/4}}{j_{0n}^2},
 \label{eq:ck_function}\\[3pt]
 \left.\partial_L f\right|_{\mathrm{lc}}&=\frac{2\xi(q_{\mathrm{lc}})f_{\mathrm{lc}}}{q_{\mathrm{lc}}L_{\mathrm{lc}}}.
 \label{eq:capture_gradient}
\end{align}
Both current components are taken from the boundary-layer solution \cite{Merritt:2015vxa} and transformed from $(-E_*/\Mstar,L^2/L_{\mathrm{c}}^2)$ \cite{Merritt:2015vxa}. Since $L_{\mathrm{lc}}$ is constant in $(E_*,L)$, the encounter capture flux per unit total energy is $F_{\mathrm{cap}}=-F_{L,\mathrm{enc}}(E_*,L_{\mathrm{lc}},t)$; orbital GW and ionization outflows are added separately.

\bibliographystyle{apsrev4-2}
\bibliography{references}
\end{document}